# NMRProcFlow: A graphical and interactive tool dedicated to 1D spectra processing for NMR-based metabolomics

D. Jacob (✉), C. Deborde, M. Lefebvre, M. Maucourt, A. Moing

UMR1332 Fruit Biology and Pathology, INRA, Univ. Bordeaux,
Plateforme Métabolome Bordeaux-MetaboHUB,
71 avenue Edouard Bourlaux, 33140 Villenave d'Ornon, France
E-mail: daniel.jacob@inra.fr

Corresponding author: **Daniel Jacob,** daniel.jacob@inra.fr

**Abstract**

**Introduction:** Concerning NMR-based metabolomics, 1D spectra processing often requires an expert eye for disentangling the intertwined peaks.

**Objectives:** Finding the best way to assist the expert in this task without requirement of programming skills.

**Methods:** NMRProcFlow was developed to be a graphical and interactive 1D NMR ($^1$H & $^{13}$C) spectra processing tool.

**Results:** NMRProcFlow (http://nmrprocflow.org), dedicated to metabolic fingerprinting and targeted metabolomic, covers all spectra processing steps including baseline correction, chemical shift calibration and alignment.

**Conclusion:** Biologists and NMR spectroscopists can easily interact and develop synergies by visualizing the NMR spectra along with their corresponding experimental-factor levels, thus setting a bridge between experimental design and subsequent statistical analyses.

## Introduction

To face the growing volumes of data generated by high-throughput analytical techniques involved in OMICS approaches, the current trend is to process these data as much as possible automatically by putting know-how and expertise into toolboxes within Virtual Research Environment (VRE), thus allowing non-experts to handle their data themselves (Candela et al. 2013). The metabolomics approach is therefore much concerned (http://phenomenal-h2020.eu/home/). Regarding metabolomics based on 1D NMR spectroscopy and although it has become a common approach, multiple challenges in spectra processing remain to be solved. As mentioned in Larive et al. (2014), continued improvement and development of new techniques for NMR data processing is essential for improving the high-throughput NMR metabolomics. As discussed in Vu et al. (2013) and Alonso et al. (2015), many of the spectra processing steps especially the spectra alignment require a certain degree of user expertise. Indeed, NMR analysis is not exempt of difficulties, and in particular subtle differences in pH, ionic strength, temperature, protein content, etc., between samples may cause differences in the NMR-detected peak position and line widths of a given metabolite (Cloarec et al. 2005; Cruz et al. 2014; Tredwell et al. 2016). In addition, in complex spectra there is a high degree of overlap in certain regions of the NMR spectrum, which hampers analysis. Therefore, the diversity of issues regarding the NMR spectra processing is not lacking, namely whose encountered during the various stages of processing (baseline correction, chemical shift calibration, removal of solvents and other contaminants, re-alignment of areas having uncontrolled variations in chemical shifts between spectra, ...) and depending on the biological context (humans, plants, micro-organisms), the type of sample source (tissue, tissue extract or biofluid ...), the analytical protocol (choice of NMR sequence, use of additives for calibration and / or quantification, use of buffer solution to stabilize pH, etc.). Given the nature of the 1D NMR spectra, and the issues cited above the expert eye is often required and even crucial to disentangle the intertwined peaks and to unravel the metabolite composition of a complex mixture sample. Apart for very well-mastered and very reproducible use cases, the implementation of 1D NMR spectra processing workflows into a toolbox and operating automatically in batch mode (regarded as a black-box) in order to be widely used by non-expert users or new-comers has not yet reached full maturity. Therefore so far the best way is still to proceed interactively with an 1D NMR spectra viewer. To fulfill this need, we have been developing NMRProcFlow, an open source software that greatly helps spectra processing (Fig. 1). It was built by involving NMR spectroscopists eager to have a quick and easy tool to use and even for non-expert users. Although an open source software such as R (R Development Core Team 2005), can be leveraged to develop a comprehensive 1D NMR spectra processing, its command-line interface is a genuine barrier for metabolomics researchers with little or no programming background. Knowing that the processing of a set of NMR spectra may be long and tedious, the purpose of NMRProcFlow is to assist users in processing 1D NMR spectra by leveraging of an easy and intuitive Graphical Unit Interface (GUI). Thus, it offers users significant time saving, while allowing them to apply their expertise without the skill barrier in programming.

## Implementation

NMRProcFlow has been developed as two separate applications (NMRspec and NMRviewer), each of them embedded in Docker images, sharing a data volume and communicating together through AJAX web services (Fig. 2). The Docker (https://www.docker.com/) technology was chosen to facilitate the setting up of the software.

NMRspec, the NMR spectra processing application has been implemented using mainly the open source software language R (R Core Team 2013, http://www.R-project.org/), and some processing algorithms have been implemented in C++ using the Rcpp package (Eddelbuettel and Francois 2011) within an internal module called Rnmr1D. The NMRspec GUI has been implemented using the 'R shiny' (http://shiny.rstudio.com/) framework based on event-driven programming. NMRviewer is based on a client-server architecture. To obtain fast transfers from server-side to client-side, first spectral data are not stored as text format but only as binary format, because text to binary conversion is time consuming. Then, the open-source software Gnuplot (http://gnuplot.info) was chosen to generate images in PNG format directly from the binary data. Thus, data reading/writing and data transfers are minimized. The NMRProcFlow tool is available online as a web tool to provide advanced processing tools for 1D NMR spectral data for anyone who needs to process NMR spectra sets, expert or not, in order to save users the trouble of installing the software. In addition, virtual appliances of NMRProcFlow compliant with the two major virtualization platforms (VMware http://www.vmware.com/ and Oracle VirtualBox https://www.virtualbox.org/) are available for a local setup thus ensuring security of confidential data such as clinical data.

**Results and Discussion**

One of the strongest point of NMRProcFlow is the possibility of visualizing the experimental factor levels within the NMR spectra set through the spectral viewer making the tool valuable to create links between the experimental design and subsequent statistical analyses, and thus facilitate interactions between biologists and NMR spectroscopists. The *NMR spectra viewer* is the central tool of NMRProcFlow and the core of the application. It allows users to visually explore the spectra overlaid or stacked, to zoom in on intensity scale, to group sets of spectra by coloring them based on their factor levels (Supplementary Information S1). Compared with the existing software based on a GUI dedicated to 1D NMR spectra processing (https://omictools.com/data-processing-category, Alonso et al. 2015), besides that some are commercial software (or implying to buy a software's license), neither of them allows specifying the factor levels within the spectra viewer, nor capturing areas of ppm to be treated either on a full set of spectra, or on a subset belonging to the same factorial group as readily as NMRProcFlow can do. Moreover, another key point linked to the ability to proceed interactively with the NMR spectra viewer within NMRProcFlow, is to allow users choosing the processing method (Fig. 1) that seems most relevant to them depending on the ppm region. For example, to align a spectral region with intertwined peaks, a Parametric Time Warping (Bloemberg et al. 2010) method will produce a better result provided that all spectra have the same amount of peaks within this region. Otherwise, a Least-Square algorithm will be more adapted (Supplementary Information S2). Similarly to the visualization, the peak alignment within a spectral region can be proceeded on either a full set of spectra or on a subset belonging to the same factor level, thus expanding the range of processing options.

The current version of NMRProcFlow accepts only pre-processed raw spectra in Bruker format, but we have planned to support more input format in a next version, in particular the 'nmrML' format (http://nmrml.org/) that should become in the near future the new NMR format standard to overcome the diversity of the vendor formats within applications (Rocca-Serra et al. 2016). "Pre-processed" means that Fourier transform and phase correction have been applied on all spectra. At this day, facing the diversity of sample sources and analytical protocols, no implementation of the automatic phasing task seems to us satisfactory without a manual readjustment. In addition, software provided by NMR manufacturers, such as TopSpin (Bruker GmbH) for example, are designed to accomplish this task, knowing that it is often performed by NMR spectroscopists as a first level of quality control. Moreover, it implies that all NMR spectra were acquired and pre-proceeded with the same number of points. However, it is possible for example to mix spectra acquired with different pulse sequences in order to compare them. Beyond its capabilities of spectra viewing and processing, NMRProcFlow is especially dedicated to metabolomics. The two major metabolomics approaches, namely metabolic fingerprinting (MF) and targeted metabolomics (TM) are taken into account. The workflow covers all steps from the spectral data up to the output data matrix (Fig. 1).

Regarding the TM approach, the identity of the metabolites of interest is established before statistical data analysis, and this involves to be able to: i) identify the spectral regions for which the quantification will be performed based on both knowledge and well-established metabolomic profiles, ii) ensure that each of these regions is not polluted by peak contribution of neighbor regions. To fulfill these two points, it is necessary to locally correct the baseline in order to i) eliminate the residual effects due to the presence of macromolecules in extracts, ii) but also reduce the prevalence of an intense peak on the less intense neighbor ones (Supplementary Information S3). The best way in the targeted metabolomics approach for obtaining buckets is to choose yourself the ppm ranges you want to integrate. Here only a few dozen of peaks corresponding to targeted and selected compounds. Their size is depending on the signal pattern. After the processing and bucketing steps, NMRProcFlow allows users to export all the data needed for the quantification into a same spreadsheet workbook. The 'qHNMR' template aggregates information within five separate tabs like the sample table, the bucket table, the Signal-to-Noise Ratio (SNR) matrix and the data matrix i.e. the values of integration for each bucket (columns) and for each spectrum (rows), and also includes another tab with the pre-calculated quantifications (Bharti et al. 2012) from data provided in the others tabs. Some information are set by default in both 'samples' and 'buckets' tabs. Just adjust them with the appropriate values and the quantifications within the eponymous tab will be automatically updated (Supplementary Information S4).

Regarding the MF approach, the identity of the metabolites of interest is established after statistical data analysis of metabolic fingerprints, and this involves to be able to: i) highlight that spectral regions having a difference between the groups are statistically significant, ii) ensure that each of these regions involves only a single metabolite peak or pattern, i.e. there is unique correspondence between a bucket and a resonance pattern (spectral signature of a metabolite). The standard approach in NMR-based metabolomics used to imply the division of spectra into equally sized bins, thereby simplifying subsequent data analysis. Yet, disadvantages are the loss of information and the occurrence of artifacts caused by peak shifts (up to 0.05 ppm Cloarec et al. 2005). Therefore, we implemented the Adaptive Intelligent Binning (De Meyer et al. 2008) algorithm which largely circumvents these problems by recursively identifying bin edges in existing bins. The later algorithm requires only minimal user input, and avoids the use of arbitrary parameters or reference spectra. It is well adapted to meet the second point mentioned above (Supplementary Information S5). Moreover, whatever the bucketing approach used, the Signal-to-Noise ratio is a good quality indicator (Supplementary Information S6). Thus, in NMRProcFlow it is possible to filter buckets based on this ratio either during the computing process, or at the output step. This possibility was carried on in a previous collaborative work where NMRProcFlow has been used with success (Bornet et al. 2016). Because statistical analysis is an important step within the MF approach, file manipulations have to be minimized to avoid laborious and time consuming conversions. Thus, NMRProcFlow can export data matrices fully compatible with online statistical analysis tools such as BioStatFlow (http://biostatflow.org) or MetaboAnalyst 3.0 (Xia et al. 2015).

In addition, NMRProcFlow allows experts to build their own spectra processing workflow, in order to become re-applicable to similar NMR spectra sets, i.e. stated as use-cases. By extension, the implementation of NMR spectra processing workflows executed in batch mode can be considered as relevant provided that we want to process in this way very well-mastered and very reproducible use cases, i.e. by applying the same Standard Operating Procedures (SOP) (Supplementary Information S7). A subset of NMR spectra is firstly processed in interactive mode in order to build a well-suited workflow. This mode can be considered as the 'expert mode'. Then, other subsets that are regarded as either similar or being included in the same case study, can be processed in batch mode, operating directly a Command Line Tool (CLI) or embedded in a workflow management system such as Workflow4Metabolomics (Giacomoni et al. 2015) or PhenoMeNal VRE App Library (http://portal.phenomenal-h2020.eu/app-library).

**Acknowledgment**The authors are grateful to the NMR users of Bordeaux Metabolome Facility for assistance in beta-testing, troubleshooting and fruitful discussions, especially E Martineau and P Giraudeau (Ceisam, Nantes), as well as the NMR members of the French-speaking Metabolomic and Fluxomic Network, especially F Fauvelle (IRMaGE, Grenoble).

**Compliance with ethical standards**

**Funding**This work was funded by INRA and MetaboHUB, the French National Infrastructure in Metabolomics and Fluxomics (ANR-11-INBS-0010 grant).

**Conflict of interest** The authors declare that they have no competing financial interests.

**Ethical approval**This article does not contain any studies with human participants or animals performed by any of the authors.

**References**

Alonso, A., Marsa, S., & Julià, A. (2015). Analytical Methods in Untargeted Metabolomics: State of the Art in 2015. *Frontiers in Bioengineering and Biotechnology*, 3, 23, doi:10.3389/fbioe.2015.00023
Bharti, S. K., & Roy, R. (2012). Quantitative 1H NMR, spectroscopy. *Trends in Analytical Chemistry* 35, 5-26, doi:10.1016/j.trac.2012.02.007
Bloemberg, T.G., Gerretzen, J., Wouters, H.J.P., Gloerich, J., van Dael, M., Wessels, H.J.C.T., et al. (2010). Improved parametric time warping for proteomics. *Chemometrics and Intelligent Laboratory Systems*, 104(1), 65-74.
Bornet, A., Maucourt, M., Deborde, C., Jacob, D., Milani, J., Vuichoud, B., et al. (2016). Highly Repeatable Dissolution Dynamic Nuclear Polarization for Heteronuclear NMR Metabolomics. *Analytical Chemistry*, 88(12), 6179–6183.
Candela, L., Castelli, D., & Pagano, P. (2013). Virtual Research Environments: an overview and a research agenda. *Data Science Journal*, 12, doi:10.2481/dsj.GRDI-013

Cloarec, O., Dumas, M., Craig, A., Barton, R., Trygg, J., Hudson, J., et al. (2005). Statistical total correlation spectroscopy: an exploratory approach for latent biomarker identification from metabolic 1H NMR data sets. *Analytical Chemistry*, 77(5), 1282-1289.

Cruz, T., Balayssac, S., Gilard, V., Martino, R., Vincent, C., Pariente, et al. (2014). 1H NMR analysis of cerebrospinal fluid from Alzheimer's disease patients: an example of a possible misinterpretation due to non-adjustment of pH. *Metabolites*, 4(1), 115-128; doi:10.3390/metabo4010115

De Meyer, T., Sinnaeve, D., Gasse, B., Tsiporkova, E., Rietzschel, E., De Buyzere, M., et al. (2008). NMR-Based Characterization of Metabolic Alterations in Hypertension Using an Adaptive, Intelligent Binning Algorithm. *Analytical Chemistry*, 80(10), 3783–3790.

Eddelbuettel, D., & Francois, R. (2011). Rcpp: Seamless R and C++ Integration. *Journal of Statistical Software*, 40(8), 1-18.

Giacomoni, F., Le Corguillé, G., Monsoor, M., Landi, M., Pericard, P., Pétéra, M.,et al. (2015). Workflow4Metabolomics: a collaborative research infrastructure for computational metabolomics. *Bioinformatics*, 31, 1493-1495

Larive, C., Barding, G. Jr, &Dinges, M. (2015). NMR Spectroscopy for Metabolomics and Metabolic Profiling. *Analytical Chemistry*, 205(87), 133-146.

Rocca-Serra, P., Salek, R.M., Arita, M., Correa, E., Dayalan, S., Gonzalez-Beltran, A., et al. (2016) *Metabolomics*, 12,14. doi:10.1007/s11306-015-0879-3

Tredwell, G.D., Bundy J.G., De Iorio M., & Ebbels, T.M.D. (2016). Modelling the acid/base $^1$H NMR chemical shift limits of metabolites in human urine. *Metabolomics*, 12,152. doi:10.1007/s11306-016-1101-y

Vu, T. N., & Laukens, K. (2013). Getting Your Peaks in Line: A Review of Alignment Methods for NMR Spectral Data. *Metabolites*, 3(2), 259–276.

Xia, J., Sinelnikov, I.V., Han, B., & Wishart, D.S. (2015). MetaboAnalyst 3.0—making metabolomics more meaningful. *Nucleic Acids Research*, 43(W1),W251-W257.doi:10.1093/nar/gkv380

**Figure Captions**

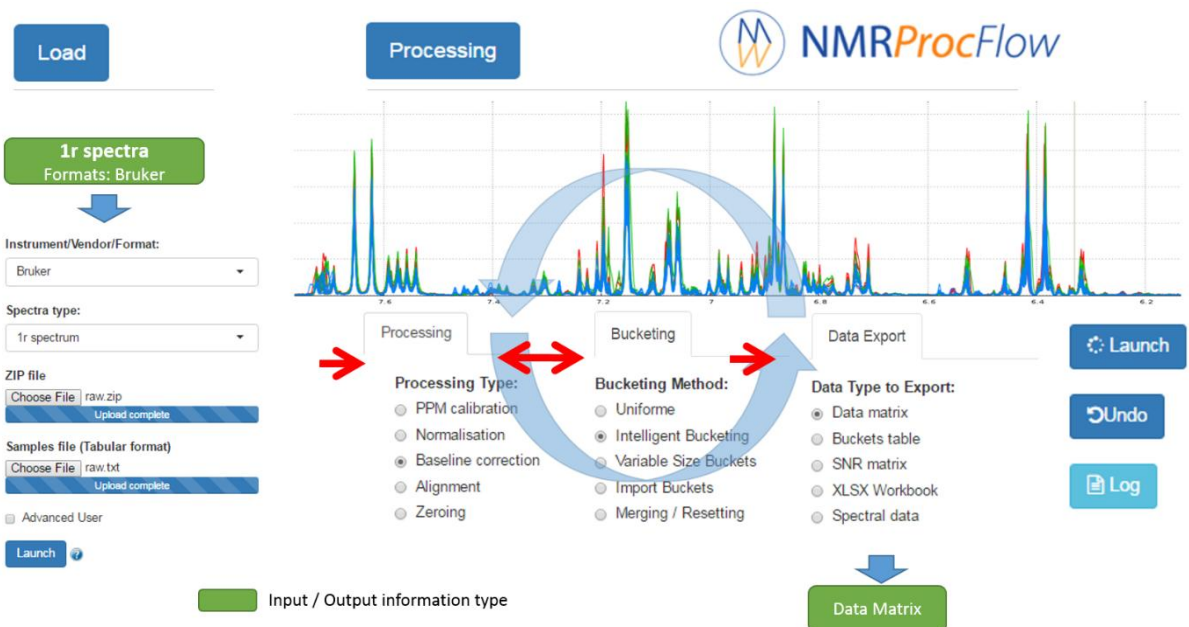

**Fig. 1** Workflow of NMRProcFlow open source software for 1D NMR spectra processing within an interactive interface based on spectra visualization.

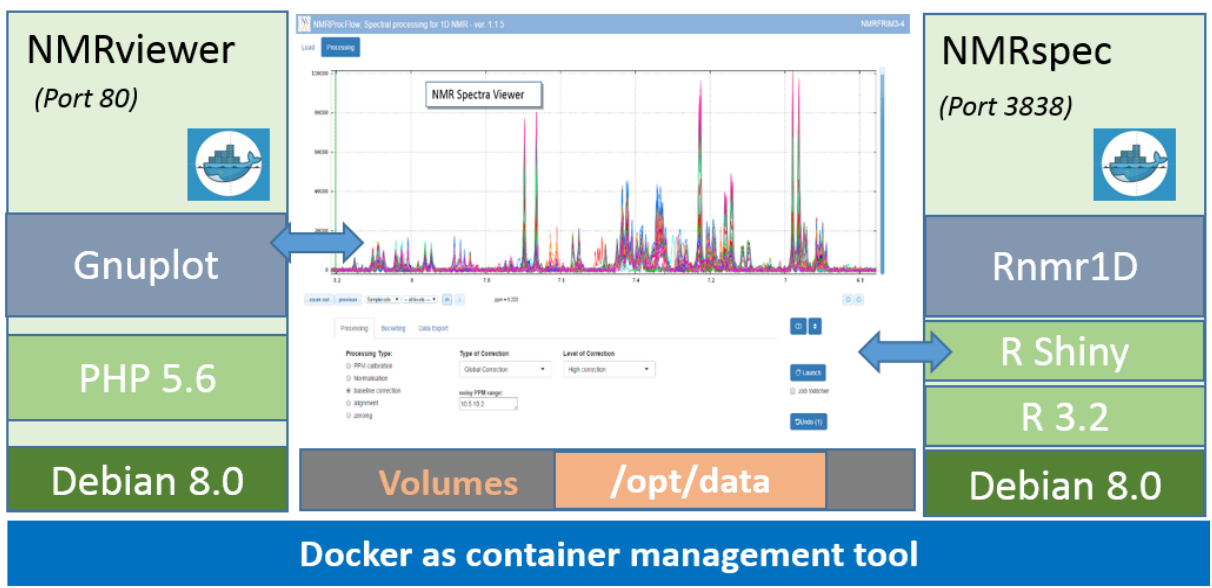

**Fig. 2** Schema of the NMRProcFlow implementation into two applications.

**Supplementary Information**

NMRProcFlow software, documentation and examples are available online at http://nmrprocflow.org.

The online documentation of the NMRProcFlow software provides a detailed description of its features and useful information on many issues not detailed in the article. This online help may advantageously be used as supplementary information. In order to not overload the manuscript with countless links, while having quick access to the desired item, the list of shortcuts mentioned in the article is provided below:

**S1**: Spectra viewer : http://nmrprocflow.org/c2 and online test of the spectra viewer: http://nmrprocflow.org/testviewer

**S2**: Spectra alignment. Example of comparison between the Least-Square and Parametric Time Warping methods http://nmrprocflow.org/c31#tabs1-3

**S3**: Baseline correction. Example of global and local baseline corrections http://nmrprocflow.org/c31#tabs1-2 and http://nmrprocflow.org/a3

**S4**: Targeted metabolomic. Quantification spreadsheet workbook http://nmrprocflow.org/a3 and http://nmrprocflow.org/themes/pdf/Targeted.pdf

**S5**: Intelligent Bucketing and Metabolomic fingerprinting http://nmrprocflow.org/a2 and http://nmrprocflow.org/c32#tabs2-1

**S6**: Signal-to-Noise Ratio (SNR) http://nmrprocflow.org/c4#tabs3-3 and http://nmrprocflow.org/themes/pdf/SNR_export.pdf

**S7**: Replaying a processing workflow in batch mode execution http://nmrprocflow.org/c6